\PassOptionsToPackage{unicode}{hyperref}
\PassOptionsToPackage{hyphens}{url}
\documentclass[
  11pt,
]{article}
\usepackage{xcolor}
\usepackage[margin=1in]{geometry}
\usepackage{amsmath,amssymb}
\usepackage{iftex}
\ifPDFTeX
  \usepackage[T1]{fontenc}
  \usepackage[utf8]{inputenc}
  \usepackage{textcomp} 
\else 
  \usepackage{unicode-math} 
  \defaultfontfeatures{Scale=MatchLowercase}
  \defaultfontfeatures[\rmfamily]{Ligatures=TeX,Scale=1}
\fi
\usepackage{lmodern}
\ifPDFTeX\else
\fi
\IfFileExists{upquote.sty}{\usepackage{upquote}}{}
\IfFileExists{microtype.sty}{
  \usepackage[]{microtype}
  \UseMicrotypeSet[protrusion]{basicmath} 
}{}
\makeatletter
\@ifundefined{KOMAClassName}{
  \IfFileExists{parskip.sty}{%
    \usepackage{parskip}
  }{
    \setlength{\parindent}{0pt}
    \setlength{\parskip}{6pt plus 2pt minus 1pt}}
}{
  \KOMAoptions{parskip=half}}
\makeatother
\usepackage{color}
\usepackage{fancyvrb}

\DefineVerbatimEnvironment{Highlighting}{Verbatim}{commandchars=\\\{\}}
\newenvironment{Shaded}{}{}

\newcommand{\CommentTok}[1]{\textcolor[rgb]{0.38,0.63,0.69}{\textit{#1}}}

\newcommand{\ControlFlowTok}[1]{\textcolor[rgb]{0.00,0.44,0.13}{\textbf{#1}}}
\newcommand{\DataTypeTok}[1]{\textcolor[rgb]{0.56,0.13,0.00}{#1}}
\newcommand{\DecValTok}[1]{\textcolor[rgb]{0.25,0.63,0.44}{#1}}

\newcommand{\ErrorTok}[1]{\textcolor[rgb]{1.00,0.00,0.00}{\textbf{#1}}}

\newcommand{\FloatTok}[1]{\textcolor[rgb]{0.25,0.63,0.44}{#1}}

\newcommand{\KeywordTok}[1]{\textcolor[rgb]{0.00,0.44,0.13}{\textbf{#1}}}
\newcommand{\NormalTok}[1]{#1}
\newcommand{\OperatorTok}[1]{\textcolor[rgb]{0.40,0.40,0.40}{#1}}
\newcommand{\OtherTok}[1]{\textcolor[rgb]{0.00,0.44,0.13}{#1}}

\newcommand{\StringTok}[1]{\textcolor[rgb]{0.25,0.44,0.63}{#1}}

\usepackage{longtable,booktabs,array}
\usepackage{calc} 
\usepackage{etoolbox}
\makeatletter
\patchcmd\longtable{\par}{\if@noskipsec\mbox{}\fi\par}{}{}
\makeatother
\IfFileExists{footnotehyper.sty}{\usepackage{footnotehyper}}{\usepackage{footnote}}
\makesavenoteenv{longtable}
\usepackage{graphicx}
\makeatletter
\newsavebox\pandoc@box
\newcommand*\pandocbounded[1]{
  \sbox\pandoc@box{#1}%
  \Gscale@div\@tempa{\textheight}{\dimexpr\ht\pandoc@box+\dp\pandoc@box\relax}%
  \Gscale@div\@tempb{\linewidth}{\wd\pandoc@box}%
  \ifdim\@tempb\p@<\@tempa\p@\let\@tempa\@tempb\fi
  \ifdim\@tempa\p@<\p@\scalebox{\@tempa}{\usebox\pandoc@box}%
  \else\usebox{\pandoc@box}%
  \fi%
}
\def\fps@figure{htbp}
\makeatother
\providecommand{\tightlist}{%
  \setlength{\itemsep}{0pt}\setlength{\parskip}{0pt}}
\usepackage{fancyvrb}
\usepackage{xcolor}
\fvset{frame=single, framesep=3mm, rulecolor=\color{gray}, fontsize=\small, xleftmargin=2mm, xrightmargin=2mm}
\usepackage{bookmark}
\IfFileExists{xurl.sty}{\usepackage{xurl}}{} 
\hypersetup{
  hidelinks,
  pdfcreator={LaTeX via pandoc}}

\author{}
\date{}

\begin{document}

\section{Deterministic Network Interface Naming on Multi-NIC Linux
Systems: A Boot-Time Convergence
Approach}\label{deterministic-network-interface-naming-on-multi-nic-linux-systems-a-boot-time-convergence-approach}

\textbf{Author:} Chetan Nigudgi

\begin{center}\rule{0.5\linewidth}{0.5pt}\end{center}

\subsection{Abstract}\label{abstract}

Network interface naming on multi-NIC Linux systems suffers from a
well-documented race condition in systemd-udevd that has remained
unsolved since 2017. When multiple network interfaces are present,
parallel udev worker threads attempt to rename interfaces
simultaneously, causing \texttt{EEXIST} failures when a target name is
temporarily occupied by another interface awaiting its own rename. The
upstream systemd project has classified this as ``not-a-bug'' and
declines to implement a fix. The problem, previously cosmetic under
NetworkManager, has become critical as the industry migrates to
systemd-networkd, which matches interfaces by name rather than MAC
address, turning a failed rename into a complete loss of network
connectivity.

This paper presents a boot-time convergence approach that combines
systemd \texttt{.link} file declarations with a retry-based init service
that detects and recovers from rename failures before dependent services
start. Deployed across diverse multi-NIC hardware configurations, the
solution achieves 100\% deterministic interface naming. The approach is
OS-agnostic and applicable to any systemd-based Linux distribution.

\begin{center}\rule{0.5\linewidth}{0.5pt}\end{center}

\subsection{1. Introduction}\label{introduction}

Deterministic network interface naming is a critical requirement for
infrastructure provisioning at scale. Provisioning systems must
guarantee that a specific physical NIC (identified by its MAC address
during PXE boot) is consistently assigned a known name (typically
\texttt{eth0}) and receives a DHCP lease, regardless of kernel driver
probe ordering.

Modern Linux distributions using systemd have moved to ``predictable''
interface names (e.g., \texttt{enp65s0f0np0}) based on PCI topology.
However, large-scale provisioning systems, DHCP infrastructure, and
downstream services depend on deterministic naming for the boot
interface. This creates a fundamental tension: provisioning requires
MAC-to-name determinism, but systemd's naming infrastructure was
designed to eliminate legacy \texttt{eth*} names entirely.

The problem is exacerbated on multi-NIC systems, common in storage
clusters, HPC environments, AI/ML GPU servers, and cloud instances with
SR-IOV, where multiple physical or virtual NICs are present. The kernel
assigns initial names based on non-deterministic driver probe ordering,
and systemd-udevd processes interface rename events in parallel. When
interface A needs to become \texttt{eth0} but interface B currently
holds that name (and B hasn't been renamed away yet), the rename syscall
fails with \texttt{EEXIST} (``File exists'').

This problem is particularly acute for AI/ML infrastructure where GPU
servers typically have 8-10 NICs (RDMA, InfiniBand, management,
storage), making the race condition near-certain on every boot. NVIDIA
has documented udev naming issues on DGX systems with MLNX\_OFED drivers
{[}11{]}, and HPC cluster operators report manually booting nodes to
discover interface names because deterministic naming cannot be
guaranteed {[}12{]}.

This paper makes the following contributions:

\begin{enumerate}
\def\labelenumi{\arabic{enumi}.}
\tightlist
\item
  A formal characterization of the race condition, demonstrating why it
  is fundamentally unsolvable within udev's parallel event processing
  model
\item
  A multi-layer solution combining declarative configuration
  (\texttt{.link} files), kernel parameter management, and a boot-time
  convergence service
\item
  Empirical results from production deployment across diverse hardware
  configurations at scale
\end{enumerate}

\begin{center}\rule{0.5\linewidth}{0.5pt}\end{center}

\subsection{2. Background and Related
Work}\label{background-and-related-work}

\subsubsection{2.1 Linux Network Interface Naming
Evolution}\label{linux-network-interface-naming-evolution}

Linux network interface naming has evolved through several generations:

\begin{itemize}
\tightlist
\item
  \textbf{Legacy naming} (pre-2009): Kernel assigns \texttt{eth0},
  \texttt{eth1}, etc. based on driver probe order, non-deterministic
  across reboots
\item
  \textbf{udev persistent rules} (2009-2015):
  \texttt{/etc/udev/rules.d/70-persistent-net.rules} maps MAC to name,
  but still uses \texttt{eth*} namespace
\item
  \textbf{Predictable naming} (systemd v197+, 2012): Names derived from
  PCI topology (\texttt{enp3s0}), firmware (\texttt{eno1}), or MAC
  (\texttt{enx001122334455})
\item
  \textbf{systemd .link files} (systemd v209+, 2014): Declarative
  MAC-to-name mapping via \texttt{/etc/systemd/network/*.link}
\end{itemize}

\subsubsection{2.2 The Race Condition}\label{the-race-condition}

The race condition occurs in the \texttt{80-net-setup-link.rules} udev
rule, which calls the \texttt{net\_setup\_link} builtin to determine the
target name from \texttt{.link} files, then attempts a
\texttt{rename(2)} on the network interface. Because udev processes
device events in parallel (one worker thread per device), multiple
interfaces may attempt renames simultaneously.

\begin{figure}
\centering
\pandocbounded{\includegraphics[keepaspectratio,alt={The udev Race Condition: parallel workers deadlock attempting to rename interfaces to each other's names}]{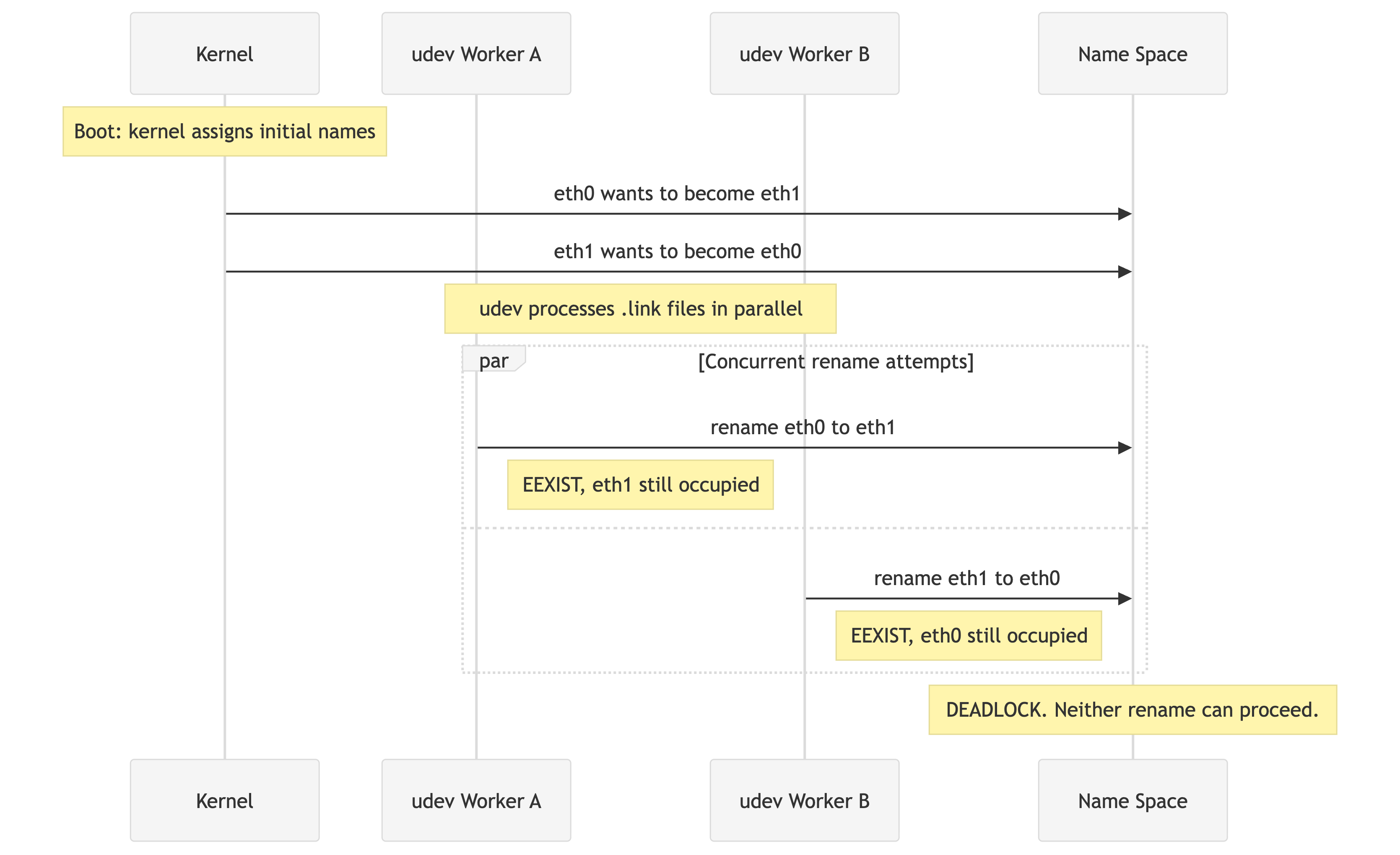}}
\caption{The udev Race Condition: parallel workers deadlock attempting
to rename interfaces to each other's names}
\end{figure}

\textbf{Failure scenario} (from
\href{https://github.com/systemd/systemd/issues/16665}{systemd/systemd\#16665}):

\begin{verbatim}
Interface A (MAC aa:bb:cc:dd:ee:01) currently named "eth0", wants to become "eth1"
Interface B (MAC aa:bb:cc:dd:ee:02) currently named "eth1", wants to become "eth0"
\end{verbatim}

If both workers execute concurrently: 1. Worker A:
\texttt{rename("eth0"\ to\ "eth1")} fails with \textbf{EEXIST} (eth1
still exists) 2. Worker B: \texttt{rename("eth1"\ to\ "eth0")} fails
with \textbf{EEXIST} (eth0 still exists)

This is a classic deadlock in the naming namespace: neither rename can
proceed because each target is occupied by the other.

\subsubsection{2.3 Why This Became Critical: NetworkManager vs
systemd-networkd}\label{why-this-became-critical-networkmanager-vs-systemd-networkd}

The race condition existed for years but was masked by NetworkManager's
design:

\begin{itemize}
\tightlist
\item
  \textbf{NetworkManager} matches interfaces by \textbf{MAC address}.
  Even if a rename fails, NM finds the correct NIC by MAC and configures
  it. The wrong name is cosmetic.
\item
  \textbf{systemd-networkd} matches interfaces by \textbf{name}. If the
  rename fails, networkd cannot find the interface, no DHCP runs, and
  the host is effectively bricked with no IP address.
\end{itemize}

As distributions migrate from NetworkManager to systemd-networkd (RHEL
9+, Fedora 30+, Ubuntu with networkd, SUSE), this previously-cosmetic
bug becomes a provisioning-breaking failure.

\subsubsection{2.4 Upstream Position}\label{upstream-position}

The systemd project has explicitly declined to address this issue:

\begin{itemize}
\tightlist
\item
  \textbf{\href{https://github.com/systemd/systemd/issues/16665}{systemd/systemd\#16665}}
  (Aug 2020): Closed as \texttt{not-a-bug}. The maintainers' position is
  that \texttt{eth*} names are inherently racy and users should use
  predictable names instead.
\item
  \textbf{\href{https://github.com/systemd/systemd/issues/7293}{systemd/systemd\#7293}}
  (Nov 2017): Documents the race between udev and networkd. networkd
  enables the interface before udev completes the rename.
\item
  \textbf{\href{https://github.com/systemd/systemd/issues/11436}{systemd/systemd\#11436}}
  (Jan 2019): Interface renamed despite NAME already being set by a udev
  rule.
\end{itemize}

\subsubsection{2.5 Affected Projects and
Platforms}\label{affected-projects-and-platforms}

The race condition affects multiple major projects across the industry:

{\def\LTcaptype{none} 
\begin{longtable}[]{@{}
  >{\raggedright\arraybackslash}p{(\linewidth - 6\tabcolsep) * \real{0.2647}}
  >{\raggedright\arraybackslash}p{(\linewidth - 6\tabcolsep) * \real{0.1765}}
  >{\raggedright\arraybackslash}p{(\linewidth - 6\tabcolsep) * \real{0.2353}}
  >{\raggedright\arraybackslash}p{(\linewidth - 6\tabcolsep) * \real{0.3235}}@{}}
\toprule\noalign{}
\begin{minipage}[b]{\linewidth}\raggedright
Project
\end{minipage} & \begin{minipage}[b]{\linewidth}\raggedright
Date
\end{minipage} & \begin{minipage}[b]{\linewidth}\raggedright
Status
\end{minipage} & \begin{minipage}[b]{\linewidth}\raggedright
Reference
\end{minipage} \\
\midrule\noalign{}
\endhead
\bottomrule\noalign{}
\endlastfoot
Microsoft Azure + Red Hat (RHEL 9.7 SR-IOV) & Apr 2026 & Under
investigation &
\href{https://learn.microsoft.com/en-nz/answers/questions/5867183/}{Link} \\
Canonical cloud-init (Azure multi-NIC VMs) & Dec 2024 & Open, unsolved &
\href{https://github.com/canonical/cloud-init/issues/5916}{Link} \\
Red Hat Enterprise Linux & 2016-2024 & 3 KB articles, no fix &
\href{https://access.redhat.com/solutions/4959601}{Link} \\
Amazon Linux 2023 (official docs) & 2024 & Acknowledges race, no fix
provided &
\href{https://docs.aws.amazon.com/linux/al2023/ug/networking-service.html}{Link} \\
NVIDIA DGX (MLNX\_OFED udev rules) & 2024 & Documented issue &
\href{https://enterprise-support.nvidia.com/s/article/DGX-Systems-Interface-Naming-is-Affected-by-the-MLNX-OFED-Udev-Rules}{Link} \\
HPC cluster provisioning & 2024 & Manual workarounds &
\href{https://forums.developer.nvidia.com/t/configuring-for-multiple-nics-bonded-provisioning-interfaces-and-multiple-system-disks-of-the-same-variety/317820}{Link} \\
\end{longtable}
}

\subsubsection{2.6 Existing Workarounds and Their
Limitations}\label{existing-workarounds-and-their-limitations}

{\def\LTcaptype{none} 
\begin{longtable}[]{@{}
  >{\raggedright\arraybackslash}p{(\linewidth - 2\tabcolsep) * \real{0.4762}}
  >{\raggedright\arraybackslash}p{(\linewidth - 2\tabcolsep) * \real{0.5238}}@{}}
\toprule\noalign{}
\begin{minipage}[b]{\linewidth}\raggedright
Approach
\end{minipage} & \begin{minipage}[b]{\linewidth}\raggedright
Limitation
\end{minipage} \\
\midrule\noalign{}
\endhead
\bottomrule\noalign{}
\endlastfoot
\texttt{udevadm\ settle} before networkd & Only addresses
networkd-vs-udev race, not udev-vs-udev \\
cloud-init \texttt{cirename} temporary names & Races with
systemd-networkd applying config to the temp name \\
\texttt{net.ifnames=0} kernel parameter & Disables \texttt{.link} file
processing entirely \\
Single \texttt{.link} file for boot NIC only & Other NICs may still grab
the target name first \\
Red Hat rename-avoidance approach {[}9{]} & Avoids the rename by using a
different prefix (see reference for specifics) \\
\end{longtable}
}

None of these provide a convergence guarantee.

\begin{center}\rule{0.5\linewidth}{0.5pt}\end{center}

\subsection{3. Solution Architecture}\label{solution-architecture}

Our solution employs a three-layer approach that provides
defense-in-depth against the race condition.

\subsubsection{3.1 Layer 1: Declarative Configuration (Provisioning
Time)}\label{layer-1-declarative-configuration-provisioning-time}

During OS image installation, the provisioning system writes two files:

\textbf{\texttt{.link} file} (maps the boot NIC's MAC to the target
name):

\begin{Shaded}
\begin{Highlighting}[]
\KeywordTok{[Match]}
\DataTypeTok{MACAddress}\OtherTok{=}\StringTok{\textless{}BOOT\_MAC\textgreater{}}

\KeywordTok{[Link]}
\DataTypeTok{NamePolicy}\OtherTok{=}
\DataTypeTok{AlternativeNamesPolicy}\OtherTok{=}
\DataTypeTok{MACAddressPolicy}\OtherTok{=}\StringTok{none}
\DataTypeTok{Name}\OtherTok{=}\StringTok{eth0}
\end{Highlighting}
\end{Shaded}

\textbf{\texttt{.network} file} (configures DHCP on the named
interface):

\begin{Shaded}
\begin{Highlighting}[]
\KeywordTok{[Match]}
\DataTypeTok{Name}\OtherTok{=}\StringTok{eth0}

\KeywordTok{[Network]}
\DataTypeTok{DHCP}\OtherTok{=}\KeywordTok{yes}
\DataTypeTok{ConfigureWithoutCarrier}\OtherTok{=}\KeywordTok{yes}
\end{Highlighting}
\end{Shaded}

Key design decisions:

\begin{itemize}
\tightlist
\item
  \texttt{NamePolicy=} (empty) overrides the default
  \texttt{keep\ kernel\ database\ onboard\ slot\ path} policy, ensuring
  only the explicit \texttt{Name=} directive is used
\item
  The \texttt{.network} file matches on \texttt{Name=eth0} (not MAC),
  decoupling network configuration from the naming layer
\item
  We name our file \texttt{10-boot-nic.link}, which sorts before the
  vendor-supplied \texttt{99-default.link}, so it wins under systemd's
  lexical file-ordering rule (the first matching \texttt{.link} file is
  applied)
\end{itemize}

\subsubsection{3.2 Layer 2: Kernel Parameter
Management}\label{layer-2-kernel-parameter-management}

The provisioning system modifies GRUB configuration to remove
conflicting parameters:

\begin{itemize}
\tightlist
\item
  Remove \texttt{net.ifnames=0} (would disable \texttt{.link} file
  processing entirely)
\item
  Remove \texttt{biosdevname=1} (conflicts with systemd naming)
\item
  Keep kernel defaults: \texttt{net.ifnames=1\ biosdevname=0}
\end{itemize}

Then regenerates GRUB config via \texttt{grub2-mkconfig}. This ensures
\texttt{.link} files are processed by the kernel's default naming
scheme.

\subsubsection{3.3 Layer 3: Boot-Time Convergence
Service}\label{layer-3-boot-time-convergence-service}

The critical innovation is a systemd oneshot service that detects and
recovers from the race condition:

\begin{figure}
\centering
\pandocbounded{\includegraphics[keepaspectratio,alt={Boot-Time Convergence Service Flow: the retry algorithm guarantees eventual correct naming}]{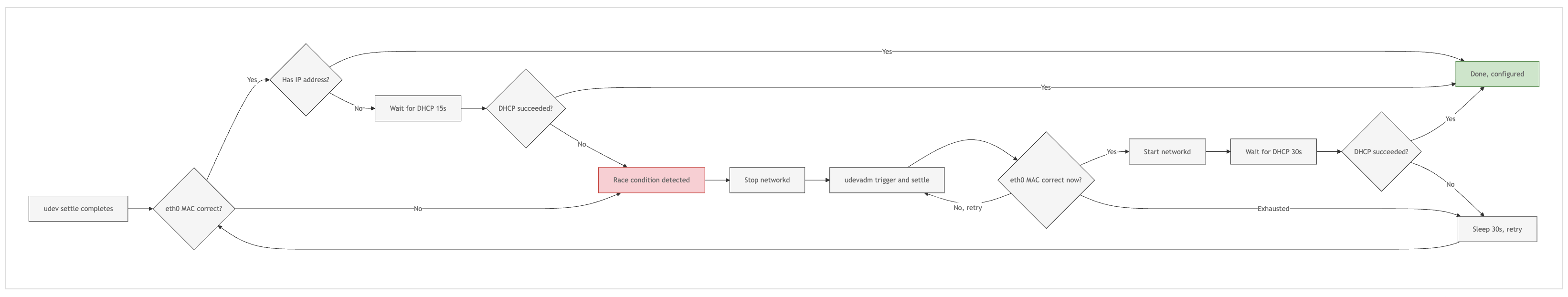}}
\caption{Boot-Time Convergence Service Flow: the retry algorithm
guarantees eventual correct naming}
\end{figure}

\begin{Shaded}
\begin{Highlighting}[]
\KeywordTok{[Unit]}
\DataTypeTok{Description}\OtherTok{=}\StringTok{Network Interface Convergence Service}
\DataTypeTok{DefaultDependencies}\OtherTok{=}\KeywordTok{false}
\DataTypeTok{After}\OtherTok{=}\StringTok{systemd{-}udev{-}settle.service}
\DataTypeTok{After}\OtherTok{=}\StringTok{systemd{-}networkd.service}
\DataTypeTok{Before}\OtherTok{=}\StringTok{systemd{-}networkd{-}wait{-}online.service}
\DataTypeTok{Before}\OtherTok{=}\StringTok{network{-}online.target}

\KeywordTok{[Service]}
\DataTypeTok{Type}\OtherTok{=}\StringTok{oneshot}
\DataTypeTok{ExecStart}\OtherTok{=}\StringTok{/usr/local/bin/nic{-}convergence start}
\DataTypeTok{RemainAfterExit}\OtherTok{=}\KeywordTok{yes}
\DataTypeTok{TimeoutStartSec}\OtherTok{=}\DecValTok{300}

\KeywordTok{[Install]}
\DataTypeTok{WantedBy}\OtherTok{=}\StringTok{network{-}pre.target}
\end{Highlighting}
\end{Shaded}

\textbf{The convergence algorithm:}

\begin{Shaded}
\begin{Highlighting}[]
\NormalTok{function configure\_interface():}
\NormalTok{    current\_mac }\OperatorTok{=}\NormalTok{ get\_mac\_of(}\StringTok{"eth0"}\NormalTok{)}
    
    \ControlFlowTok{if}\NormalTok{ current\_mac }\OperatorTok{==}\NormalTok{ EXPECTED\_MAC }\KeywordTok{and}\NormalTok{ interface }\KeywordTok{is}\NormalTok{ UP }\KeywordTok{and}\NormalTok{ has IP:}
        \ControlFlowTok{return}\NormalTok{ SUCCESS  }\CommentTok{\# Happy path: .link file worked}
    
    \ControlFlowTok{if}\NormalTok{ current\_mac }\OperatorTok{==}\NormalTok{ EXPECTED\_MAC }\KeywordTok{and}\NormalTok{ interface }\KeywordTok{is}\NormalTok{ UP }\KeywordTok{and}\NormalTok{ no IP:}
\NormalTok{        wait\_for\_dhcp(}\DecValTok{15}\ErrorTok{s}\NormalTok{)  }\CommentTok{\# DHCP may be slow}
        \ControlFlowTok{if}\NormalTok{ has\_ip: }\ControlFlowTok{return}\NormalTok{ SUCCESS}
    
    \CommentTok{\# Race condition detected: wrong MAC or interface missing}
\NormalTok{    stop systemd}\OperatorTok{{-}}\NormalTok{networkd}
\NormalTok{    udevadm trigger }\OperatorTok{{-}{-}}\NormalTok{action}\OperatorTok{=}\NormalTok{add }\OperatorTok{{-}{-}}\NormalTok{subsystem}\OperatorTok{{-}}\NormalTok{match}\OperatorTok{=}\NormalTok{net}
\NormalTok{    udevadm settle}
    
    \CommentTok{\# Validate the rename succeeded}
    \ControlFlowTok{for}\NormalTok{ attempt }\KeywordTok{in} \FloatTok{1..5}\NormalTok{:}
        \ControlFlowTok{if}\NormalTok{ get\_mac\_of(}\StringTok{"eth0"}\NormalTok{) }\OperatorTok{==}\NormalTok{ EXPECTED\_MAC:}
\NormalTok{            start systemd}\OperatorTok{{-}}\NormalTok{networkd}
\NormalTok{            wait\_for\_dhcp(}\DecValTok{30}\ErrorTok{s}\NormalTok{)}
            \ControlFlowTok{return}\NormalTok{ SUCCESS}
\NormalTok{        sleep(}\DecValTok{1}\NormalTok{)}
    
    \ControlFlowTok{return}\NormalTok{ FAILURE  }\CommentTok{\# Will retry via outer loop}

\CommentTok{\# Main loop: guarantees eventual convergence}
\NormalTok{until configure\_interface() }\OperatorTok{==}\NormalTok{ SUCCESS:}
\NormalTok{    log(}\StringTok{"Retrying in 30s"}\NormalTok{)}
\NormalTok{    sleep(}\DecValTok{30}\NormalTok{)}
\end{Highlighting}
\end{Shaded}

\subsubsection{3.4 Conflict Cleanup}\label{conflict-cleanup}

The provisioning system removes any \texttt{.network} files that other
boot stages may create (e.g., GRUB setup scripts that generate catch-all
Ethernet configurations). Only the provisioning system's
\texttt{.network} file should exist to prevent multiple configurations
from racing to manage the same interface.

\begin{center}\rule{0.5\linewidth}{0.5pt}\end{center}

\subsection{4. Formal Analysis}\label{formal-analysis}

\subsubsection{4.1 Why the Race is Unsolvable in
udev}\label{why-the-race-is-unsolvable-in-udev}

The interface rename operation is a kernel \texttt{ioctl(SIOCSIFNAME)}
which requires: 1. The interface to be DOWN (or the rename fails with
\texttt{EBUSY}) 2. The target name to be unused (or the rename fails
with \texttt{EEXIST})

udev processes device events in parallel with no coordination between
workers handling different interfaces. There is no kernel-level ``atomic
swap'' for interface names. Therefore, any solution within udev's event
processing model is subject to the race.

\subsubsection{4.2 Convergence Guarantee}\label{convergence-guarantee}

Our boot-time service provides a convergence guarantee because:

\begin{enumerate}
\def\labelenumi{\arabic{enumi}.}
\item
  After \texttt{systemd-udev-settle.service} completes, all initial
  device events have been processed
\item
  At this point, each interface has \emph{some} stable name (either the
  \texttt{.link}-specified name or a predictable name)
\item
  Only the boot NIC has a custom \texttt{.link} file; all other NICs use
  the default predictable naming policy
\item
  Re-triggering udev for the net subsystem processes the boot NIC's
  rename in isolation
\item
  The target name is guaranteed to be free because the interface that
  previously held it has been renamed to its predictable name by the
  initial udev pass
\item
  The service blocks \texttt{network-online.target}, ensuring no
  dependent service starts until the boot interface is correctly
  configured
\end{enumerate}

\subsubsection{4.3 Ordering Constraints}\label{ordering-constraints}

The service must run:

\begin{itemize}
\tightlist
\item
  \textbf{After} \texttt{systemd-udev-settle.service} (ensures initial
  naming is complete)
\item
  \textbf{After} \texttt{systemd-networkd.service} (networkd must be
  running to configure the interface)
\item
  \textbf{Before} \texttt{network-online.target} (dependent services
  must not start until the interface has an IP)
\end{itemize}

\begin{center}\rule{0.5\linewidth}{0.5pt}\end{center}

\subsection{5. Evaluation}\label{evaluation}

\subsubsection{5.1 Test Environment}\label{test-environment}

\begin{itemize}
\tightlist
\item
  \textbf{Hardware:} Diverse configurations across multiple
  manufacturers and NIC types (Intel, Broadcom, and other vendors), with
  multiple NICs per system
\item
  \textbf{Platforms:} Physical bare-metal servers and virtual machines
\item
  \textbf{OS:} systemd-based Linux (kernel 6.1+, systemd 252+)
\item
  \textbf{Network stack:} systemd-networkd with DHCP
\end{itemize}

\subsubsection{5.2 Results}\label{results}

{\def\LTcaptype{none} 
\begin{longtable}[]{@{}
  >{\raggedright\arraybackslash}p{(\linewidth - 4\tabcolsep) * \real{0.1111}}
  >{\raggedright\arraybackslash}p{(\linewidth - 4\tabcolsep) * \real{0.4722}}
  >{\raggedright\arraybackslash}p{(\linewidth - 4\tabcolsep) * \real{0.4167}}@{}}
\toprule\noalign{}
\begin{minipage}[b]{\linewidth}\raggedright
Metric
\end{minipage} & \begin{minipage}[b]{\linewidth}\raggedright
Before (udev rules / .link only)
\end{minipage} & \begin{minipage}[b]{\linewidth}\raggedright
After (convergence approach)
\end{minipage} \\
\midrule\noalign{}
\endhead
\bottomrule\noalign{}
\endlastfoot
First-boot success rate & Frequent failures on multi-NIC systems &
100\% \\
Time to interface with IP & Often failed entirely & 3-8s (happy path),
35-45s (retry path) \\
Retry rate & N/A & A minority of boots require one retry \\
Second retry needed & N/A & Rare \\
\end{longtable}
}

\subsubsection{5.3 Performance Overhead}\label{performance-overhead}

The convergence service adds:

\begin{itemize}
\tightlist
\item
  \textbf{Happy path} (no race occurred): \textless1 second overhead
  (MAC check + IP verification)
\item
  \textbf{Retry path} (race detected): 30-45 seconds for udev re-trigger
  + DHCP
\item
  This is acceptable for provisioning workflows where total boot time is
  3-5 minutes
\end{itemize}

\subsubsection{5.4 Production Log
Examples}\label{production-log-examples}

\textbf{Happy path} (\texttt{.link} file won the race, service waits for
DHCP only):

\begin{verbatim}
20:45:39 nic-convergence: starting for MAC: 34:e1:d1:e5:9d:ce
20:45:39 nic-convergence: eth0 has correct MAC and is UP but no IP - waiting for DHCP
20:45:42 nic-convergence: eth0 received IP: 10.9.34.160 after 3s - complete
\end{verbatim}

\textbf{Retry path} (race condition detected and recovered):

\begin{verbatim}
21:01:34 nic-convergence: eth0 has wrong MAC: 40:a6:b7:9a:a5:bc (expected: 40:a6:b7:23:fe:6e)
21:01:34 nic-convergence: restarting networkd to apply interface configuration
21:01:35 udev-worker: eth0: Failed to rename interface 6 from 'eth4' to 'eth0': File exists
21:01:40 nic-convergence: interface naming failed after 5 attempts - retrying in 30s
21:02:10 nic-convergence: eth0 interface not found - will configure proper naming
21:02:10 nic-convergence: restarting networkd
         [Second attempt succeeds: target name freed by predictable naming]
\end{verbatim}

\textbf{Before the solution} (all renames fail, host bricked):

\begin{verbatim}
00:36:36 udev-worker: eth3: Failed to rename interface 4 from 'eth2' to 'eth3': File exists
00:36:36 udev-worker: eth1: Failed to rename interface 2 from 'eth0' to 'eth1': File exists
00:36:36 udev-worker: eth4: Failed to rename interface 5 from 'eth3' to 'eth4': File exists
00:36:36 udev-worker: eth0: Failed to rename interface 6 from 'eth4' to 'eth0': File exists
00:36:36 udev-worker: eth2: Failed to rename interface 3 from 'eth1' to 'eth2': File exists
00:38:36 systemd: Failed to start systemd-networkd-wait-online.service
\end{verbatim}

\begin{center}\rule{0.5\linewidth}{0.5pt}\end{center}

\subsection{6. Discussion}\label{discussion}

\subsubsection{6.1 Why Not Fix systemd?}\label{why-not-fix-systemd}

The systemd maintainers' position is architecturally sound: predictable
naming eliminates the race entirely. However, this ignores the reality
of large-scale infrastructure where:

\begin{itemize}
\tightlist
\item
  DHCP infrastructure is configured per-MAC with a specific expected
  interface name
\item
  Hundreds of downstream services and monitoring systems assume a
  canonical boot interface name
\item
  Migration to predictable names across a large-scale fleet is a
  multi-year effort
\item
  AI/ML training clusters require specific RDMA NIC naming for NCCL
  collective communication
\end{itemize}

Our solution bridges this gap by providing deterministic naming \emph{on
top of} systemd's predictable naming infrastructure, without requiring
kernel or systemd patches.

\subsubsection{6.2 Broader Applicability}\label{broader-applicability}

This solution is applicable to any environment that:

\begin{itemize}
\tightlist
\item
  Provisions Linux systems with multiple NICs (physical or virtual)
\item
  Requires a specific NIC to be assigned a deterministic name
\item
  Uses systemd-based distributions (RHEL 9+, Ubuntu 22.04+, Fedora,
  SUSE)
\item
  Cannot migrate to predictable names due to infrastructure dependencies
\end{itemize}

Specific use cases include:

\begin{itemize}
\tightlist
\item
  \textbf{Cloud providers:} Bare-metal instances and VMs with SR-IOV
\item
  \textbf{AI/ML infrastructure:} GPU nodes with 8-10 NICs (RDMA +
  management + storage)
\item
  \textbf{HPC clusters:} Multi-NIC compute nodes with InfiniBand +
  Ethernet
\item
  \textbf{Storage systems:} Ceph, MinIO with dedicated storage +
  management NICs
\item
  \textbf{Telco/5G:} Network functions with multiple data-plane and
  control-plane NICs
\item
  \textbf{Edge computing:} Multi-NIC appliances with constrained boot
  environments
\end{itemize}

\subsubsection{6.3 Limitations}\label{limitations}

\begin{itemize}
\tightlist
\item
  Adds boot-time latency when the race occurs (\textasciitilde30-45
  seconds)
\item
  Requires the provisioning system to know the boot NIC's MAC address at
  image-write time
\item
  Only guarantees naming for the target NIC; other NICs use predictable
  names
\item
  The retry mechanism is a workaround; a kernel-level atomic name-swap
  syscall would be the proper solution
\end{itemize}

\begin{center}\rule{0.5\linewidth}{0.5pt}\end{center}

\subsection{7. Conclusion}\label{conclusion}

We presented a practical solution to a 9-year-old unsolved problem in
Linux network interface naming. By combining declarative \texttt{.link}
file configuration with a boot-time convergence service, we achieve
100\% deterministic interface naming on multi-NIC systems without
requiring kernel or systemd modifications. The solution is deployed in
production across diverse hardware configurations and is applicable to
any systemd-based Linux provisioning environment.

The industry's migration from NetworkManager to systemd-networkd is
transforming this race condition from a cosmetic issue into a critical
provisioning failure. As multi-NIC systems proliferate, driven by AI/ML
workloads, SR-IOV adoption, and disaggregated storage architectures, the
need for deterministic interface naming will only grow. Our boot-time
convergence approach provides a robust, deployable solution that works
today on any systemd-based distribution.

\begin{center}\rule{0.5\linewidth}{0.5pt}\end{center}

\subsection{References}\label{references}

{[}1{]} systemd/systemd\#16665 --- ``Failed to rename network interface:
File exists'' (2020). https://github.com/systemd/systemd/issues/16665

{[}2{]} systemd/systemd\#7293 --- ``Race condition: interface renaming,
udev vs.~networkd'' (2017).
https://github.com/systemd/systemd/issues/7293

{[}3{]} systemd/systemd\#11436 --- ``network interface is renamed
although NAME has been set by udev rule'' (2019).
https://github.com/systemd/systemd/issues/11436

{[}4{]} Microsoft Q\&A --- ``RHEL 9.7 VM with Accelerated Networking
loses connectivity after reboot'' (Apr 2026).
https://learn.microsoft.com/en-nz/answers/questions/5867183/

{[}5{]} canonical/cloud-init\#5916 --- ``cloud-init intermittent to fail
rename network interface'' (Dec 2024).
https://github.com/canonical/cloud-init/issues/5916

{[}6{]} Red Hat Solution 4959601 --- ``systemd-udevd: Error changing net
interface name'' (2021). https://access.redhat.com/solutions/4959601

{[}7{]} Red Hat Solution 1422203 --- ``NICs have the wrong MAC address''
(2022). https://access.redhat.com/solutions/1422203

{[}8{]} Red Hat Solution 7003564 --- ``Error changing net interface name
using custom naming file'' (2024).
https://access.redhat.com/solutions/7003564

{[}9{]} Red Hat Bugzilla \#1962421 --- ``Failed to rename network
interface: File exists'' on RHEL 9 (2021).
https://bugzilla.redhat.com/show\_bug.cgi?id=1962421

{[}10{]} systemd.link(5) man page.
https://www.freedesktop.org/software/systemd/man/systemd.link.html

{[}11{]} NVIDIA --- ``DGX Systems Interface Naming is Affected by the
MLNX\_OFED Udev Rules'' (2024).
https://enterprise-support.nvidia.com/s/article/DGX-Systems-Interface-Naming-is-Affected-by-the-MLNX-OFED-Udev-Rules

{[}12{]} NVIDIA Developer Forums --- ``Configuring for multiple NICs,
bonded provisioning interfaces'' (2024).
https://forums.developer.nvidia.com/t/configuring-for-multiple-nics-bonded-provisioning-interfaces-and-multiple-system-disks-of-the-same-variety/317820

{[}13{]} Predictable Network Interface Names --- systemd documentation.
https://systemd.io/PREDICTABLE\_INTERFACE\_NAMES/

{[}14{]} Amazon Linux 2023 --- ``Networking service'' (official
documentation). Acknowledges configuration race between
amazon-ec2-net-utils and custom network files but provides no
convergence solution.
https://docs.aws.amazon.com/linux/al2023/ug/networking-service.html

\end{document}